\documentclass[11pt]{article}

\newsavebox{\junk}
\savebox{\junk}[1.6mm]{\hbox{$|\!|\!|$}}

\def\bbbz{{\mathchoice {\hbox{$\sf\textstyle Z\kern-0.4em Z$}}
{\hbox{$\sf\textstyle Z\kern-0.4em Z$}}
{\hbox{$\sf\scriptstyle Z\kern-0.3em Z$}}
{\hbox{$\sf\scriptscriptstyle Z\kern-0.2em Z$}}}}
\def\sq{\hbox{\rlap{$\sqcap$}$\sqcup$}}

\def\bbbr{{\rm I\!R}}

\usepackage{epsfig}
\usepackage{amsfonts}
\usepackage{latexsym}
\usepackage{bm}
\usepackage[scanall]{psfrag}

\newcommand{\ben}{\begin{enumerate}}
\newcommand{\een}{\end{enumerate}}
\newcommand{\bit}{\begin{itemize}}
\newcommand{\eit}{\end{itemize}}

\newtheorem{theorem}{Theorem}[section]
\newtheorem{proposition}[theorem]{Proposition}

\newcommand{\ba}{\begin{array}{rcl}}
\newcommand{\ea}{\end{array}}
\newcommand{\bt}{\begin{theorem}}
\newcommand{\et}{\end{theorem}}
\newcommand{\bd}{\begin{description}}
\newcommand{\ed}{\end{description}}

\def\slabel#1{\label{s:#1}}

\def\elabel#1{\label{e:#1}}
\def\eq#1/{(\ref{e:#1})}

\def\Section#1{Section~\ref{s:#1}}
\def\Re{\bbbr}

\def\state{{\sf X}}

\def\beq{\begin{equation}}
\def\eeq{\end{equation}}
\def\beqa{\begin{eqnarray}}
\def\eeqa{\end{eqnarray}}

\def\qed{\ifmmode\sq\else{\unskip\nobreak\hfil
\penalty50\hskip1em\null\nobreak\hfil\sq
\parfillskip=0pt\finalhyphendemerits=0\endgraf}\fi}

\def\sqr#1#2{{\vcenter{\hrule height.#2pt
      \hbox{\vrule width.#2pt height#1pt \kern#1pt
         \vrule width.#2pt}
       \hrule height.#2pt}}}

\newcommand{\bc}{\begin{corollary}}
\newcommand{\ec}{\end{corollary}}

\newcommand{\bp}{\begin{proposition}}
\newcommand{\ep}{\end{proposition}}

\def\eye(#1){{\bf (#1)}\quad}

\def\taboo#1{{{}_{#1}}}

\def\0P{\taboo{0}P}
\def\0Pn{\taboo{0}P^n}

\date{}

\title{A Class Coupler for Perfect Sampling from
Continuous Distributions With and Without Atoms}

\def\jem{Postal Address:
      Department of Applied Mathematics,
      University of Colorado, Box 526
      Boulder CO 80309-0526, USA; phone: 303-492-0685; email: mao@colorado.edu, corcoran@colorado.edu}

\author {W. Mao and J.N. Corcoran\\
University of Colorado \thanks{\jem} }

\begin{document}


\maketitle
\vspace{-.8cm}

\begin{abstract} \small \noindent
We consider the simulation of distributions that are a mixture of
discrete and continuous components. We extend a Metropolis-Hastings-based perfect sampling
algorithm of Corcoran and Tweedie \cite{cortwe01b} to allow for a
broader class of transition candidate densities. The resulting
algorithm, know as a ``class coupler'', is fast to implement and is applicable to purely discrete or purely continuous
densities as well. Our work is motivated by the study of a composite
hypothesis test in a Bayesian setting via posterior simulation and
we give simulation results for some problems in this area.

\bigskip

\end{abstract}

\footnotetext{Keywords: {MCMC, perfect simulation, Metropolis-Hastings, 
Bayesian hypothesis testing}
}

\setcounter{page}{0}



\section{Introduction}
\slabel{int}
It has been more than a decade since the appearance of the seminal
paper 
of Propp and Wilson \cite{prowil96} which introduced perfect simulation to the
Monte Carlo community. Immediately thereafter, several variations and
extensions
\cite{fil98,fostwe97a,fostwecor98,hagetal96,ken98,mol99,murgre97}
appeared, proving to be effective in areas such as statistical physics,
 spatial
point processes and operations research, where they
provided simple and powerful
alternatives to  existing  methods based on, for example,  iterating
transition laws.

In this paper, we extend a perfect sampling algorithm of Corcoran and
Tweedie \cite{cortwe01b} so that it is applicable to a larger class of
probability 
models. In particular, our algorithm is useful 
for simulating densities that are a mixture of continuous and discrete
densities.
Our approach was motivated by the study of a composite hypothesis test in a
Bayesian setting such as

$$
H_{0} : \theta = \theta_{0}
$$
versus
$$
H_{1} : \theta \ne \theta_{0},
$$
where $\theta$ is a model parameter with a ``mixed'' distribution that
takes on values in a continuum but allows for the possibility that
$\theta = \theta_{0}$. Thus, it becomes necessary to impose a mixed prior
density on $\theta$ which results in a mixed posterior density as well.

In order to simulate values from the mixed posterior
density we have adapted a perfect version of the
Metropolis-Hastings algorithm in \cite{cortwe01b} where we somewhat relax an
assumption that a Markov chain candidate transition density proposes
future states independent of current states.
We refer to our algorithm as a ``class coupler'' as it relies on a
partitioning of the state space into subspaces or classes. In the
context of the above described mixture target density the atom at
$\theta_{0}$ is designated as one (single point) class while the
remaining possibilities for $\theta$ comprise the members of a second
class. In general though, we may decompose the state space of the
target density into several classes and it is not necessary that there
be any atoms.

In \Section{perfect}, we give a brief review of the concept of perfect
sampling, the Metropolis-Hastings algorithm,  and of the existing
perfect 
version of the
Metropolis-Hastings algorithm of Corcoran and Tweedie
\cite{cortwe01b}. In \Section{classcoupler}, we describe our extension
which is known as the class coupler. In \Section{examples}, we provide
some simulation examples for Bayesian regression models and compare
our results to those of Gottardo and Raftery \cite{gotraft2004}.

\section{Perfect Sampling}
\slabel{perfect}
A Markov chain Monte Carlo (MCMC) algorithm that would enable one to
draw values from a given density $\pi(\cdot)$ such as
is a recipe for creating a Markov chain $\{X_{n}\}$
so that
\beq
\elabel{mc}
\lim_{n \rightarrow \infty} P(X_{n} \in A) = \pi(A).
\eeq
Here, $A$ is a subset of the state space of $\{X_{n}\}$ and $\pi(A)$ is
shorthand notation for $\int_{A} \pi(x)\, dx$. 

Quantities involving $\pi(\cdot)$ are then usually approximated by
simulating values of $X_{N}$ for some very large $N$.  The advantage of
{\bf perfect sampling} (also known as {\bf{perfect simulation}}) over
this 
traditional MCMC approach allows us
to directly sample (simulate) values of $X_{\infty}$.  

The essential idea behind perfect sampling is to find a random epoch
$-T$ in the {\emph{past}} such that, if we construct sample paths
(according to a
transition law $P(x,A):=P(X_{n+1} \in A | X_{n} =x)$ that is converging
to $\pi$) from every point in
the state space starting at $-T$,  then all paths will come together and
meet or ``couple'' by time zero. The common value of the paths at time
zero is a
draw from $\pi$.
Intuitively, it is clear why  this result holds with such a
random time $T$ as we are constructing the tail end of a path that has
traveled forward from time $-\infty$ since any such path will pass
through the time point $-T$ and then be ``funneled forward'' to the common
value at time zero. We refer to the smallest value of $T$ for which this
can be achieved as a \emph{backward coupling time} (BCT).

For more details on perfect sampling, we refer the interested reader
to Casella, Lavine, and Robert \cite{casrob00}.

\subsection{The (Non-Perfect) Metropolis-Hastings Algorithm}
\slabel{imh}

Suppose we wish to simulate values from a target distribution with
density $\pi(x)$ for $x \in \Re^{k}$. The Metropolis-Hastings algorithm
\cite{tie94} is one way to construct a Markov chain and transition
probabilities with our target distribution as its limiting distribution.

In order to describe the Metropolis-Hastings algorithm(s), we consider a
\emph{candidate transition kernel} $Q(x, A)$ for $x \in \Re^{k}$ and a
Borel set $A \subset \Re^{k}$ satisfying
$$
Q(x,A) \geq 0 \qquad \mbox{and} \qquad Q(x,\Re^{k})=1
$$
which generates potential transitions for a discrete-time Markov chain
evolving on $\Re^{k}$. We assume that there exists a density $q(x,y)$
such that
$$
Q(x,A) = \int_{A} q(x,y) \, dy.
$$

The Metropolis algorithm \cite{metropolis}, dating back to 1953, uses a
symmetric candidate transition $Q$ for which $q(x,y) = q(y,x)$. In 1970,
Hastings \cite{Hastings} extended the Metropolis algorithm to a more
general
$Q$. In either case, the simulator proceeds by generating candidate
transitions from state $x$ to state $y$ according to the distribution $Q$,
and accepting the transition with probability
\beq
\elabel{alpha}
\alpha(x,y) = \left \{
\begin{array}{ll}
\min \left \{ 1, \,\, {\displaystyle
    \frac{\pi(y)}{\pi(x)}\frac{q(x,y)}{q(y,x)} }\right \}
& \pi(x)q(y,x) > 0 \\
1 & \pi(x)q(y,x) = 0.
\end{array}
\right.
\eeq
Thus evolves a Markov chain with transition density
$$
p(x,y) = q(x,y) \, \alpha(x,y), \qquad y \ne x,
$$
which will remain at the same point with probability
$$
P(x,\{x\}) = \int q(x,y) \, [1-\alpha(x,y)] \, dy.
$$

It is easy to verify that $\pi$ is the \emph{invariant} or
\emph{stationary}
measure for the chain in the sense that
$$
\pi(A) = \int \pi(x) P(x,A) \, dx, \qquad \forall A \in {\cal {B}}(\state)
$$
where $\state$ is the state space of the chain and ${\cal {B}}(\state)$
are the Borel sets in $\state$.

It is also easy to verify that any limiting distribution is stationary
and that, for this chain constructed with the Metropolis-Hastings
algorithm, there is a unique stationary distribution. Thus, the
stationary and limiting distributions are one and the same.  

Note that due to its presence only in ratios, we can run this algorithm
even if we only know $\pi$ up to a constant of proportionality.

In order to simulate a value drawn from $\pi$, one must generally
select a distribution $Q$ and run a Metropolis-Hasting sample path for ``a
long time'' until it is suspected that convergence to $\pi$ has been
achieved. Choices for $Q$ and rates of convergence have been studied
extensively in \cite{cai97},\cite{mentwe96}, and \cite{stramertweedie99}, for
example.

\subsection{The Perfect Independent Metropolis-Hastings Algorithm}
\slabel{cortwe}

In Corcoran and Tweedie \cite{cortwe01b}, a Metropolis-Hastings-based
perfect sampling
algorithm was introduced, eliminating the need to address issues of
convergence.

We use the term ``independent'' to
describe the Metropolis-Hastings algorithm where candidate states are
generated by a distribution $Q$ that is independent of the current state
of the chain. In this Section, we assume the existence of a density $q(x,y)$ such
that
$$
Q(x,A) = \int_{A} q(x,y) \, dy.
$$

Assuming an independent candidate density means that
$$
q(x,y) \equiv q(y).
$$  

The perfect independent Metropolis-Hastings (perfect IMH) algorithm uses
the ratios in the acceptance probabilities
given by (\ref{e:alpha}) to reorder the states in such a way that we always
accept moves to the left (or downwards). That is, if we write
$\pi(x) = k h(x)$ where $k$ is possibly unknown, we define the IMH ordering,
\beq
x \succeq y \qquad \Longleftrightarrow \qquad
\frac{\pi(y)q(x)}{\pi(x)q(y)} \geq 1
 \qquad \Longleftrightarrow \qquad \frac{h(y)}{q(y)} \geq \frac{h(x)}{q(x)}
\elabel{order1}
\eeq

With this ordering, we can (hopefully) attain a ``lowest state'' $\ell$
for which $h(l)/q(l) \geq h(x)/q(x)$ for all $x$ in the state space.
Given the role of $h/q$ in the acceptance probability of the
Metropolis-Hasting algorithm, one can think of $\ell$ as the state
that is 
hardest to move away
from when running the IMH algorithm. Thus, if we are able to accept a move
from $\ell$ to a candidate state $y$ drawn from the distribution $Q$ with
density $q$, then sample paths from every point in the state space will also
accept a move to $y$, so all possible sample paths will couple. For
the following formal description of the steps of the algorithm, we
will assume the existence of a ``highest'' point $u$ for which
$h(u)/q(u) \leq h(x)/q(x)$ for all $x$. The existence of such a point
is not required though as will be noted directly after the algorithm description.

\vspace{0.2in}
{\bf{Perfect IMH Algorithm}}

\ben
\item Draw a sequence of random variables $Q_{-n} \sim Q$ for
$n=0,1,2,\ldots$,
and a sequence $\alpha_{-n} \sim \mbox{Uniform}(0,1)$ for $n=1,2,\ldots$.
\item For each time $-n=-1,-2,\ldots$, start a lower path $L $ at $\ell$,
and an upper path, $ U$ at $u$.
\item

\ben
\item For the lower path:  Accept  a move from $\ell$ to $Q_{-n+1}$ at time
$-n+1$ with probability  $\alpha(l,Q_{-n+1})$, otherwise remain at state
$\ell$.
That is, accept the move from $\ell$ to $Q_{-n+1}$ if $\alpha_{-n} \leq
\alpha(l,Q_{-n+1})$.

\item For the upper path: Similarly, accept   a move from $u$ to
$Q_{-n+1}$ at time $-n+1$ if $\alpha_{-n} \leq \alpha(u,Q_{-n+1})$;
otherwise remain at state $Q_{-n}$.
\een

\item Continue until $T$ defined as the first  $n$ such that at time $-n+1$
each of these two paths accepts the point $Q_{-n+1}$. (Continue the
Metropolis-Hastings
algorithm forward to time zero using the same sequence of $\alpha$'s
and $Q$'s produced in steps
1-3
to get the draw from $\pi$ at time zero.)
\een

There is a monotonicity imposed by (\ref{e:order1}) such that for any
candidate point $y$,
$$
x_{1} \succeq x_{2} \qquad \Rightarrow \qquad \alpha(x_{1},y) < \alpha(x_{2},y).
$$
Consequently, if any point accepts a move to $y$, all ``higher''
points will also accept a move to this candidate.
Therefore, the upper path in this algorithm will accept a candidate point whenever the
lower path will, and the two paths will be the same from that time forward.
Consequently, our description of the upper process is a formality only, and,
indeed, the upper process need not be run at all. Therefore, the point
$u$ does not need to be identified and in fact is not even required to
exist! 

\begin{figure}[htp]
\caption{A Realization of the Perfect IMH Algorithm With Backward
  Coupling
Time at -4}
\label{fig:qprocess}
\begin{center}
\ \psfig{figure=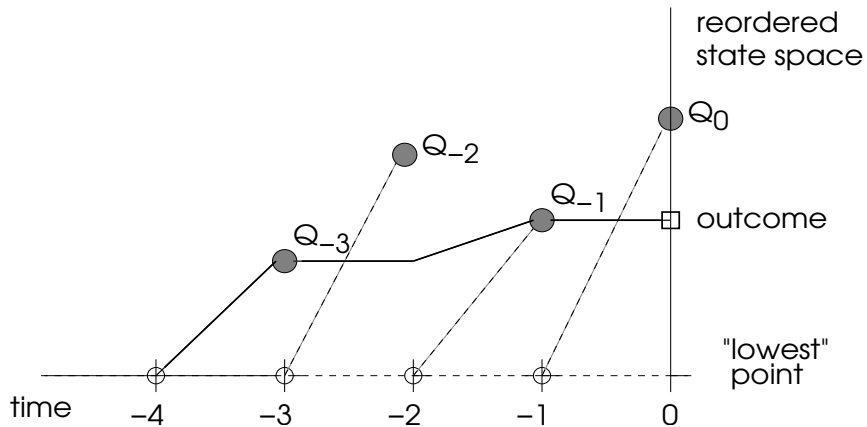,height=2.2in,width=4.5in} \\
\footnotesize{
\begin{minipage}[t]{6in}
Dashed grey lines represent potential but unrealized arcs of the
sample
path.
Solid grey lines represent the sample paths started at times -1, -2,
and -3
that did not achieve the coupling. The solid black line represents the
path
whose outcome is ultimately observed in the perfect sampling
algorithm.
\end{minipage}
}
\end{center}
\end{figure}

Figure \ref{fig:qprocess}
illustrates a realization of the perfect IMH algorithm.

\section{Our Extension: ``The Class Coupler''}
\slabel{classcoupler}

We now extend the perfect IMH algorithm to allow for a transition candidate
density that can depend on the current state of the chain up to the
inclusion of that state in a set found in a partition of the state
space. Specifically, we will partition the state space into sets or
``classes'' ${\cal{C}}_{1}, {\cal{C}}_{2}, \ldots, {\cal{C}}_{K}$ and
allow for a transition candidate density of the form
$$
q(x,y) = q_{{\cal{C}}_{i}}(y) \qquad \mbox{for} \,\,\, x \in {\cal{C}}_{i}    
$$  
for some independent transition candidate density
$q_{{\cal{C}}_{i}}(\cdot)$.

We begin with a simple two class partition.

\subsection{A Continuous Distribution With a Single Atom}
\slabel{contwithatom}
Suppose that we have a sample $\vec{X}=(X_{1}, X_{2}, \ldots, X_{m})$
from a density $f(\vec{x}|\theta)$ with a prior density $f_{\theta}(\cdot)$
of the form
\beq
\elabel{prior1}
f_{\theta} (\cdot) = p \cdot \delta_{0}(\cdot) + (1-p)
f_{1}(\cdot),
\eeq
where $f_{1}(\cdot)$ is a continuous density, $0 < p < 1$,
and $\delta_{0}(\cdot)$ is the Dirac delta function with point mass
concentrated at zero. $f_{1}$ may depend on known hyperparameters.

Further suppose that
we wish to test
$$
H_{0}  :  \theta = 0 \qquad \mbox{versus} \qquad H_{1}  :  \theta \ne 0.
$$
by drawing values from the posterior distribution with density
\beq
\elabel{pi8}
\pi(\theta) \equiv \pi(\theta | \vec{x})
\propto f(\vec{x} | \theta) \cdot f_{\theta} (\theta).
\eeq
(The null value $0$ is used here only for simplicity and may be
replaced with a generic $\theta_{0}$.)

In this paper, we are concerned only with the Monte Carlo algorithm that
will allow us to obtain perfect draws from $\pi(\theta)$ and not with what
one should do with such values in order to make a decision about the
given hypotheses. Typically, one would  report and interpret posterior odds ratio or a Bayes factor. We refer the reader to \cite{raft96} for details about Bayesian hypothesis testing in general. 


In order to simulate the density in (\ref{e:pi8}), we will use a Metropolis-Hastings algorithm with transition candidate density
\beq
\elabel{candidate}
q(\theta,\theta^{\prime}) = \left\{
\begin{array}{lcl}
\delta_{0}(\theta^{\prime}) & , & \mbox{if} \,\,\, \theta \ne 0\\
\\
g(\theta^{\prime}) & , & \mbox{if} \,\,\, \theta = 0\\
\end{array}
\right.
\eeq
where $g(\cdot)$ is some continuous density that will be chosen in a
convenient way. Note that this candidate density is no longer an
independent candidate density. That is, the right-hand side of
(\ref{e:candidate}) is not independent of $\theta$.

For the regular forward (non-perfect) Metropolis-Hastings algorithm, one
would proceed to draw approximate values from $\pi(\theta)$ as follows.  
\begin{enumerate}
\item Start with some (perhaps arbitrary) value of $\theta$.
\item Propose another value $\theta^{\prime}$ by drawing a value from the
distribution with density $q(\theta,\cdot)$.
\item With probability
$$
\alpha (\theta,\theta^{\prime}) = \min \{ 1, r(\theta,\theta^{\prime}) \}
$$
where
$$
r(\theta,\theta^{\prime}) = \frac{\pi(\theta^{\prime})
q(\theta^{\prime},\theta)}{\pi(\theta) q(\theta,\theta^{\prime})} =
\frac{f(\vec{x}|\theta^{\prime})
f_{\theta}(\theta^{\prime})
q(\theta^{\prime},\theta)}{f(\vec{x}|\theta) f_{\theta}(\theta)
q(\theta,\theta^{\prime})}.
$$
accept the move to $\theta^{\prime}$ and set $\theta= \theta^{\prime}$.

\item Return to Step 2.

\end{enumerate}

After ``many'' iterations of Steps 2 through 4, one could output the 
current value
of the chain as an approximate draw from $\pi(\theta)$. We would arrive at
a perfect draw from $\pi(\theta)$ after an infinite number of iterations.

\begin{figure}[htp]
\caption{Possible Updates of Sample Paths of the MH-Chain  with
Candidate Transitions Generated by $q(\theta,\theta^{\prime})$ from
(\ref{e:candidate})}
\label{fig:mucouple}
\begin{center}
\ \psfig{figure=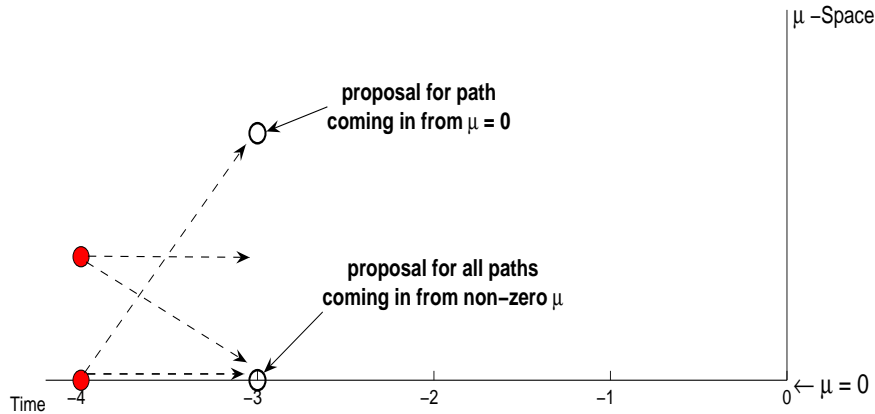,height=3.2in,width=5in} \\
\end{center}
\end{figure}

In order to turn this into a perfect simulation algorithm, we need to
run through backward time steps (taking care to reuse the random variates
associated with each time step) and we need to  be
able to figure out if and when all possible paths wandering through the
state space of $\theta$ will meet. Consider Figure \ref{fig:mucouple} which
depicts the possible updates of sample paths of the MH-chain between
time steps $-4$ and $-3$. At time $-4$ the point $\theta=0$ is
depicted as well as an arbitrary non-zero value for $\theta$.

\begin{figure}[htp]
\caption{One Case of Coupling}
\label{fig:mucouple1}
\begin{center}
\ \psfig{figure=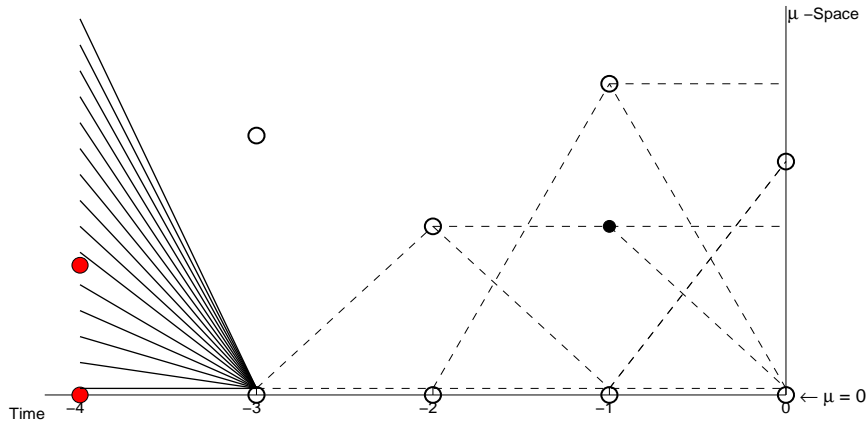,height=2.5in,width=5.0in} \\
\footnotesize{
\begin{minipage}[t]{6in}
All sample paths have coupled at time $-3$. From there, dashed lines
represent potential paths forward to time zero following a series of
possible acceptance or rejections of candidate points which are
depicted
as open circles. For this particular image, the perfect draw from
$\pi$ will be one of 4 points indicated at time zero, depending on
which dashed lines are followed.
\end{minipage}
}
\end{center}
\end{figure}

One way to achieve a coupling of all sample paths is in
the case that {\bf all} sample paths starting at $\theta \ne 0$ accept
the proposed value of $0$ {\bf and} the path starting at $\theta = 0$
rejects its
proposed non-zero value and stays at zero. This is depicted in Figure
\ref{fig:mucouple1}.

\begin{figure}[htp]
\caption{A Second Case of Coupling With a Backward Coupling Time of
  $-4$}
\label{fig:mucouple2}
\begin{center}
\ \psfig{figure=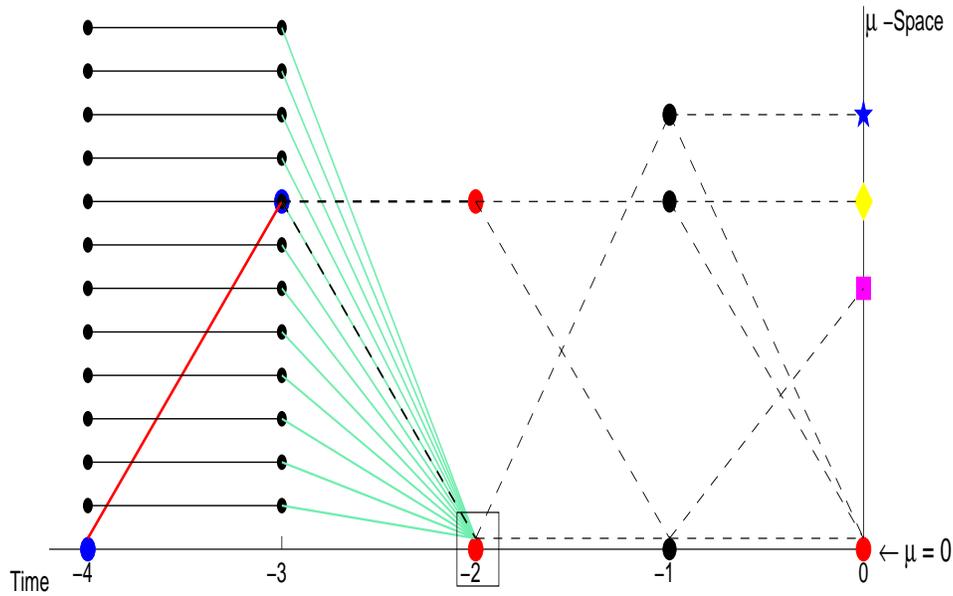,height=3.5in,width=5.5in} \\
\footnotesize{
\begin{minipage}[t]{6in}
Open circles in this Figure represent candidate values for possible
transitions. Dashed lines represent potential transitions while solid
lines represent actual realized transitions.

\vspace{0.1in}
From time $-4$ to time $-3$: All sample paths from non-zero $\theta$
have accepted the candidate
value of $0$ at time $-3$. The sample path from state $0$ has accepted
the non-zero candidate value.

\vspace{0.1in}
From time $-3$ to time $-2$: The path that is non-zero at time $-3$
fails to accept the $0$ candidate value at time $-2$ and ``stays
flat''. That path that is at $0$ at time $-3$ accepts the proposed
non-zero candidate at time $-2$.

\vspace{0.1in}
From time $-2$ to time $-1$: We are now following two non-zero
paths. In this depiction, both have accepted the $0$ candidate and
coupling is achieved at time $-1$. Continuing this single path forward
to time $0$ results in a perfect draw from the desired distribution
which is shown in a box. Assuming previously failed attempts starting
from all
points at times

$-1$, $-2$, and $-3$, we say that the backward coupling time is $-4$.

\end{minipage}
}
\end{center}
\end{figure}

Another way to achieve coupling is in the case that {\bf
  all} sample paths starting at $\theta \ne 0$ accept the proposed value
of $0$ {\bf and} the path starting at $\theta = 0$ accepts its proposed
non-zero value, but at a later time (before time $0$) the resulting
two paths manage to meet. This is depicted in Figure
\ref{fig:mucouple2}. (It is important to note that this realization
only becomes relevant after failed attempts at achieving full coupling
starting first at time $-1$, then at time $-2$, and then at time $-3$.)

To determine when {\bf all} non-zero paths accept a zero candidate, we
must find
\beq
\elabel{minprob}
\min_{\theta \ne 0} \, r(\theta,0).  
\eeq
This is the smallest probability of acceptance of zero for all non-zero
paths. In a simulation, we would have random numbers, $\alpha_{-1}, \alpha_{-2},
\ldots$, uniformly distributed on $[0,1]$, associated with each
negative time step. All non-zero paths at time $-t$ will accept a move
to the proposal zero at time $-t+1$ if
$$
\alpha_{-t} \leq \min_{\theta \ne 0} r(\theta,0).
$$

As we are free to choose the density $g(\cdot)$ in (\ref{e:candidate}),
we choose it to be the same as $f_{1}(\cdot)$, which is the
density used in the mixture prior (\ref{e:prior1}). This results in a
simplification of $r(\theta,\theta^{\prime})$.

Since
\beq
\elabel{mune0}
r(\theta,0) = \frac{p}{1-p} \cdot
\frac{f(\vec{x}|0)}{f(\vec{x}|\theta)},
\eeq
$r(\theta,0)$ is clearly minimized at the maximum likelihood estimator
(MLE) of $\theta$, which we will denote by $\hat{\theta}$.

\subsubsection{Complete Algorithm Details}
\slabel{example1}


We include this Section in order to clarify the details of the class
coupler especially as it pertains to reusing random deviates from
earlier time steps. It serves as a reference that may be useful in
algorithm implementation but can be skipped in a reading of this paper

%

\vspace{0.2in}
{\bf Algorithm:}

Let $t = 1$. Compute 
$$
r(\hat{\theta},0) = \frac{p}{1-p} \cdot\frac{f(\vec{x}|0)}{f(\vec{x}|\hat{\theta})}.
$$

\begin{enumerate}
\item Generate random deviates $Q_{-t+1} \sim
f_{1}$ and $\alpha_{-t} \sim \mathrm{Uniform}(0,1).$\\
  Compute 
$$
r(0,Q_{-t+1}) = \frac{1-p}{p} \cdot
\frac{f(\vec{x}|Q_{-t+1})}{f(\vec{x}|0)}.
$$
\item If
  \begin{itemize}
  \item $r(\hat{\theta},0) < \alpha_{-t} \le r(0,Q_{-t+1})$, we have
    achieved 
the one-step
    coupling depicted in Figure \ref{fig:mucouple1}. Set $X_{-t+1} = 0$. Record the backward coupling time as $T=t$. Go to Step $4$.
  \item $\alpha_{-t} \le \min(r(\hat{\theta},0),r(0,Q_{-t+1}))$, we have achieved the 
first
    stage of the coupling depicted in Figure \ref{fig:mucouple2}. Set 
$X_{-t+1}^{(1)} = Q_{-t+1} $ and
    $Q_{-t+1}^{(2)} = 0$. If $t>1$, go to Step $3$ to attempt the
    second 
stage of coupling. If $t=1$, return to Step 1. 
\end{itemize}
  Otherwise, set $t=t+1$ and return to Step 1.
  
\item Second stage for two stage coupling: Run the usual forward MH
simulations, using existing random deviates $Q_{-t+2}, Q_{-t+3} \ldots, Q_{0}$ and $\alpha_{-t+1}, \alpha_{-t+2} \ldots, \alpha_{1}$ until both paths, started at $X_{-t+1}^{(1)}$ and $X_{-t+1}^{(2)}$ reach time zero.

For $i=1,2$, this is accomplished by setting, for $k=t-2,
  t-3, \ldots, 0$,
$$
X_{-k}^{(i)} = \left\{
\begin{array}{lcl}
Q_{-k}&,& \mbox{if} \,\,\, \alpha_{-k-1} \leq r(X_{-k-1}^{(i)},Q_{-k})\\
\\
X_{-k-1}^{(i)} &,& \mbox{otherwise}
\end{array}
\right.
$$

  \begin{itemize}
  \item If $X_{0}^{(1)} = X_{0}^{(2)}$, stop. $X_{0}^{(1)}$ is a perfect
    draw from target distribution and the backward coupling time is
    $T=t$.
  \item If $X_{0}^{(1)} \ne X_{0}^{(2)}$, set $t=t+1$ and return to Step 1.
  \end{itemize}
  
\item Run the single path, starting from $X_{-t+1}=0$, forward using
  the existing random deviates $Q_{-t+2}, Q_{-t+3} \ldots, Q_{0}$ and
  $\alpha_{-t+2}, \alpha_{-t+3} \ldots, \alpha_{1}$ by, for $k=t-2,
  t-3, \ldots, 0$, setting
$$
X_{-k} = \left\{
\begin{array}{lcl}
Q_{-k}&,& \mbox{if} \,\,\, \alpha_{-k-1} \leq r(X_{-k-1},Q_{-k})\\
\\
X_{-k-1} &,& \mbox{otherwise}
\end{array}
\right.
$$

Stop.  The value reached at time zero, $X_{0}$, is a perfect
    draw from the target distribution.    
\end{enumerate}  

\subsection{A More General Algorithm}
We extend the perfect algorithm used in \Section{contwithatom} to a more
general situation. There, we partitioned the state space for
$\theta$ into the point $0$ and the class of non-zero points. In
general, 
we can to
partition the state space  into two or more non-overlapping classes.
 In the case of $K$ classes, we modify the algorithm described
 in \Section{contwithatom} by using a single candidate value for all
points in class $i$ from a density, $q_{i}(\cdot)$. For simplicity,
especially as it pertains to computing acceptance probabilities, it
is advisable to not include class $i$ points in the support of $q_{i}(\cdot)$. 
For example, in the case of two classes, labeled I and II, we
recommend proposing  single candidate value for all class I
points from a density, $q_{I}(\cdot)$,  with support in class II and
proposing a single candidate value for all class II
points from a density, $q_{II}(\cdot)$ with support in class I. Allowing ``within-class'' proposals may lower the backward coupling times but will complicate the minimization of acceptance probabilities.

Each time an entire class of points accepts a transition candidate
point, the cardinality of the number of sample paths to be followed is
reduced. The goal, of course, is to reduce to a single path before
time $0$. 

\section{Examples}
\slabel{examples}

\subsection{A Simple Bayesian Regression Model}
\slabel{simpleregression}
Consider the model
\beq
\elabel{model1}
Y_{j} = \mu+ \epsilon_{j}, \qquad j=1,2,\ldots,n
\eeq
where the $\epsilon_{j}$ are independent and identically distributed
as
$$
\epsilon_{j}|\sigma^{2} \sim N(0,v).
$$

We assume a normal prior on $\mu$ with an atom at $0$:
$$ 
f (\mu|p,\sigma_{\mu}^{2}) = p \cdot \delta_{0}(\cdot) + (1-p)
\cdot N(\mu;0,\sigma_{\mu}^{2} )
$$
with known hyperparameters $p$ and $\sigma_{\mu}^{2}$. Here,
$N(\cdot;0,\sigma_{\mu}^{2})$ is the normal density with mean $0$ and
variance $\sigma_{\mu}^{2}$.

We assume that we do not know the variance parameter for
$\epsilon_{j}$ but we impose an inverse gamma prior with known
hyperparameters $k_{1}$ and $k_{2}$. 
(This means that $1/v$ has
a gamma distribution with mean $k_{1}/k_{2}$ and variance
$k_{1}/k_{2}^{2}$.) We will assume that, a priori, $\mu$ and
$v$ are independent, so that the posterior (target) density has
the form
$$
\pi(\mu,v) \equiv \pi(\mu,v|\vec{y}) \propto
f(\vec{y}|\mu,v) \cdot f(\mu|p,\sigma_{\mu}^{2})
\cdot IG(v;k_{1},k_{2})
$$
where $IG(\cdot;k_{1},k_{2})$ is the inverse gamma density.


We will simulate values from $\pi(\mu,v)$ using the class
coupler with
$$
\mbox{class I} = \{(0,v): v > 0\}
$$  
and
$$
\mbox{class II} = \{(\mu,v): -\infty < \mu < \infty, \mu \ne 0,
v > 0\}.
$$

Consider the candidate transition density
$$
q((\mu,v),(\mu^{\prime},v^{\prime})) =
q_{1}(\mu,\mu^{\prime}) \cdot q_{2}(v,v^{\prime})
$$
where
$$
q_{1}(\mu,\mu^{\prime})= \left\{
\begin{array}{lcl}
\delta_{0}(\mu^{\prime}) & , & \mbox{if} \,\,\, \mu \ne 0\\
\\
g(\mu^{\prime}) = N(\mu^{\prime};0,\sigma_{\mu}^{2}) & , & \mbox{if}
\,\,\, \mu = 0\\
\end{array}
\right.
$$
and
$$
q_{2}(v,v^{\prime}) = q_{2} (v^{\prime }) =
IG(v^{\prime };k_{1},k_{2}).
$$

\vspace{0.1in}
Clearly, class II points will always have a proposed transition in class
I. In order to determine that all points accept this proposal, we need
to minimize
$$
\begin{array}{lcl}
r((\mu,v),(0,v^{\prime})) & = & {\displaystyle
\frac{f(\vec{y}|0,v^{\prime }) \cdot
f(0|p,\sigma_{\mu}^{2}) \cdot
  q((0,v^{\prime}),(\mu,v))}{f(\vec{y}|\mu,v)
  \cdot f(\mu| p,\sigma_{\mu}^{2}) \cdot
  q((\mu,v),(0,v^{\prime}))} }\\
\\[0.1in]
&=& {\displaystyle \frac{p}{1-p} \cdot
\frac{f(\vec{y}|0,v^{\prime })}{f(\vec{y}|\mu,v)} }
\end{array}
$$
over class II points. The minimum occurs at the MLEs
$\hat{\mu}$ and $\hat{v})$ where
$$
\hat{\mu} = \overline{Y}=\frac{1}{n} \sum_{i=1}^{n} Y_{i} \qquad
\mbox{and} \qquad \hat{v} = \frac{1}{n}
\sum_{i=1}^{n}(Y_{i}-\overline{Y})^2.
$$
 
\vspace{0.1in}
Similarly, class I points will always have a proposed transition in
class II. In order to determine that all points accept this proposal, we
need to minimize
$$
\begin{array}{lcl}
r((0,v),(\mu^{\prime},v^{\prime})) &=&
\frac{\displaystyle f(\vec{y}|\mu^{\prime},v^{\prime}) \cdot
f(\mu^{\prime}|p,\sigma_{\mu}^{2}) \cdot
q((\mu^{\prime},v^{\prime}),(0,v))}{\displaystyle
f(\vec{y}|0,v)
\cdot
f(0|p,\sigma_{\mu}^{2}) \cdot
q((0,v),(\mu^{\prime},v^{\prime }))}\\
\\
&=& \frac{\displaystyle 1-p}{\displaystyle p} \cdot \frac{\displaystyle
  f(\vec{y}|\mu^{\prime},v^{\prime}) }{\displaystyle
f(\vec{y}|0,v) }
\end{array}
$$
over class I points. The minimum occurs at the restricted MLE 
$$
\hat{v}_{0} = \frac{1}{n} \sum_{i=1}^{n} Y_{i}^{2}.
$$

\vspace{0.2in}
{\bf Algorithm:}

Let $t=3$. (A full coupling of sample paths will take at least three
steps due to the multi-stage nature of the coupler.) Generate and store
independent random deviates $S_{0}$, $S_{-1}$, $N_{0}$, $N_{-1}$,  and
$U_{-1}$, $U_{-2}$,  where $S_{i} \sim IG(k_{1},k_{2})$, $N_{i} \sim
N(0,\sigma_{\mu}^{2})$, and $U_{i} \sim \mbox{Uniform}(0,1)$.

\begin{enumerate}
\item Generate and store $S_{-t+1} \sim IG(k_{1},k_{2})$, $N_{-t+1} \sim
N(0,\sigma_{\mu}^{2})$, and $U_{-t} \sim \mbox{Uniform}(0,1)$.

Compute
$$
r((\hat{\mu},\hat{v}),(0,S_{-t+1})) = \frac{p}{1-p} \cdot
\frac{f(\vec{y}|0,S_{-t+1})}{f(\vec{y}|\hat{\mu},\hat{v})}
$$
and
$$
r((0,\hat{v}_{0}),(N_{-t+1},S_{-t+1})) = \frac{1-p}{p} \cdot
\frac{f(\vec{y}|N_{-t+1},S_{-t+1})}{f(\vec{y}|0
,\hat{v}_{0})}
$$
\item If
\begin{itemize}
\item $U_{-t} \leq \min (r((\hat{\mu},\hat{v}),(0,S_{-t+1})),r((0,\hat{v}_{0}),(N_{-t+1},S_{-t+1})))$, we have achieved the first
stage of the desired coupling. Set $X_{-t+1}^{(1)} = (0,S_{-t+1})$ and
$X_{-t+1}^{(2)} = (N_{-t+1},S_{-t+1})$. Set $T=t$.
\item $U_{-t} > \min (r((\hat{\mu},\hat{v},(0,S_{-t+1})),r((0,\hat{v}_{0}),(N_{-t+1},S_{-t+1})))$, set $t=t+1$ and return to
Step 1.
\end{itemize}
\item Run forward to time zero as follows.

Let $k=-t+1$, and let $X_{k}^{(i)}(j)$ denote the $j$th component of
$X_{k}^{(i)}$ for $j=1,2$.
\begin{enumerate}

\item For $i=1,2$,
\begin{itemize}
\item If $X_{k}^{(i)}(1)=0$,  compute
$$
r((X_{k}^{(i)}(1),X_{k}^{(i)}(2)),(N_{k+1},S_{k+1})) = \frac{p}{1-p} \cdot
\frac{f(\vec{y}|N_{k+1},S_{k+1})}{f(\vec{y}|(X_{k}^{(i)}(1),X_{k}^{(i)}(2)))}.
$$
If $U_{k} \leq r((X_{k}^{(i)}(1),X_{k}^{(i)}(2)),(N_{k+1},S_{k+1}))$, set $X_{k+1}^{(i)}=(N_{m+1},S_{k+1})$.

Otherwise, set $X_{k+1}^{(i)}=X_{k}^{(i)}$.
\item If $X_{k}^{(i)}(1) \ne 0$,  compute
$$
r(((X_{k}^{(i)}(1),X_{k}^{(i)}(2)),(0,S_{k+1})) = \frac{1-p}{p} \cdot
\frac{f(\vec{y}|0,S_{k+1})}{f(\vec{y}|(X_{k}^{(i)}(1),X_{k}^{(i)}(2)))}.
$$
If $U_{k} \leq r(((X_{k}^{(i)}(1),X_{k}^{(i)}(2)),(0,S_{k+1}))$, set $X_{k+1}^{(i)}=(0,S_{k+1})$. 

Otherwise, set
$X_{k+1}^{(i)}=X_{k}^{(i)}$.

\end{itemize}
\item Set $k=k+1$.
\begin{itemize}
\item If $k<0$, return to Step 3a.
\item If $k=0$ and $X_{0}^{(1)} \ne X_{0}^{(2)}$, set $t=t+1$ and
return to Step 1.
\item If $k=0$ and $X_{0}^{(1)} = X_{0}^{(2)}$, Stop. $X_{0}^{(1)}$ is a
perfect draw from $\pi(\mu,v)$ and the backward coupling time
is $T$.
\end{itemize}
\end{enumerate}
\end{enumerate}

\begin{figure}[htp]
\caption{A Two class Coupler}
\label{fig:mucouple3}
\begin{center}
\ \psfig{figure=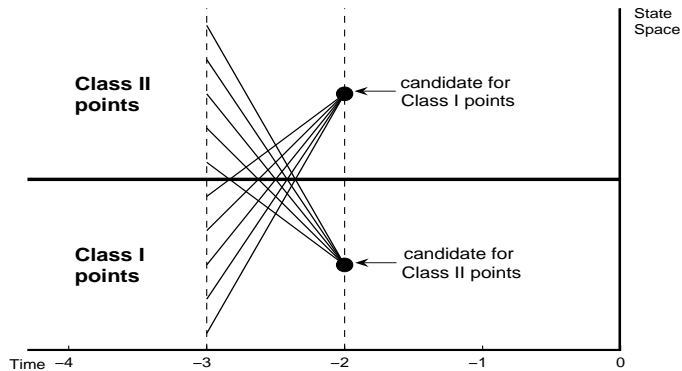,height=2.2in,width=4in} \\
\footnotesize{
\begin{minipage}[t]{6in}
Coupling {\bf may} occur in this two class case when {\bf all} paths
from class I accept their candidate value which is proposed in class
II
 and {\bf all} paths
from class II accept their candidate value which is proposed in class
I.
After this, we need to follow only two sample paths forward to time
zero. Coupling occurs {\bf if} these two paths meet before time zero.
\end{minipage}
}
\end{center}
\end{figure}

\vspace{0.2in}
It is important to note that in the first bullet point of Step 2 of
the above algorithm, we have required that all class I points accept a
candidate {\bf{and}} all class II points accept a class I candidate in
the same time step. This is depicted in Figure \ref{fig:mucouple3}.
This is a conservative algorithm in the sense that it will give a
longer backward coupling time than allowing class acceptances in
different time steps, however it is much simpler to execute. Allowing
both classes to accept in different time steps will require us to potentially
follow many more paths, the number of which will be variable and
generally increasing in time.

\vspace{0.2in}
{\bf Results: Simulation One}

For this simulation, we used data and hyperparameters identical to those
used in a standard non-perfect Monte Carlo simulation in Gottardo and
Raftery \cite{gotraft2004}. They provided $10$ randomly generated 
observations from a $N(0.5,1)$ distribution as:
$$
0.575, \,\,\, 1.808, \,\,\, 0.532, \,\,\, -0.168, \,\,\, 0.529, \,\,\,
0.888, \,\,\, -1.368, \,\,\, -0.512, \,\,\, 2.667, \,\,\, 0.874.
$$
Hyperparameters were $p=0.5$,  $\sigma_{\mu}^{2} = 100$, $k_{1}=1$, and $k_{2}=0.05$.

\begin{figure}[htp]
\caption{Simulation Results for \Section{simpleregression}, Simulation
  One a) Histogram of 100,000 Values of $\mu$ Drawn
from the Posterior $\pi(\mu,v)$, b) Histogram of
Corresponding Backward Coupling Times}
\label{fig:page11mus}
\begin{center}
\ \psfig{figure=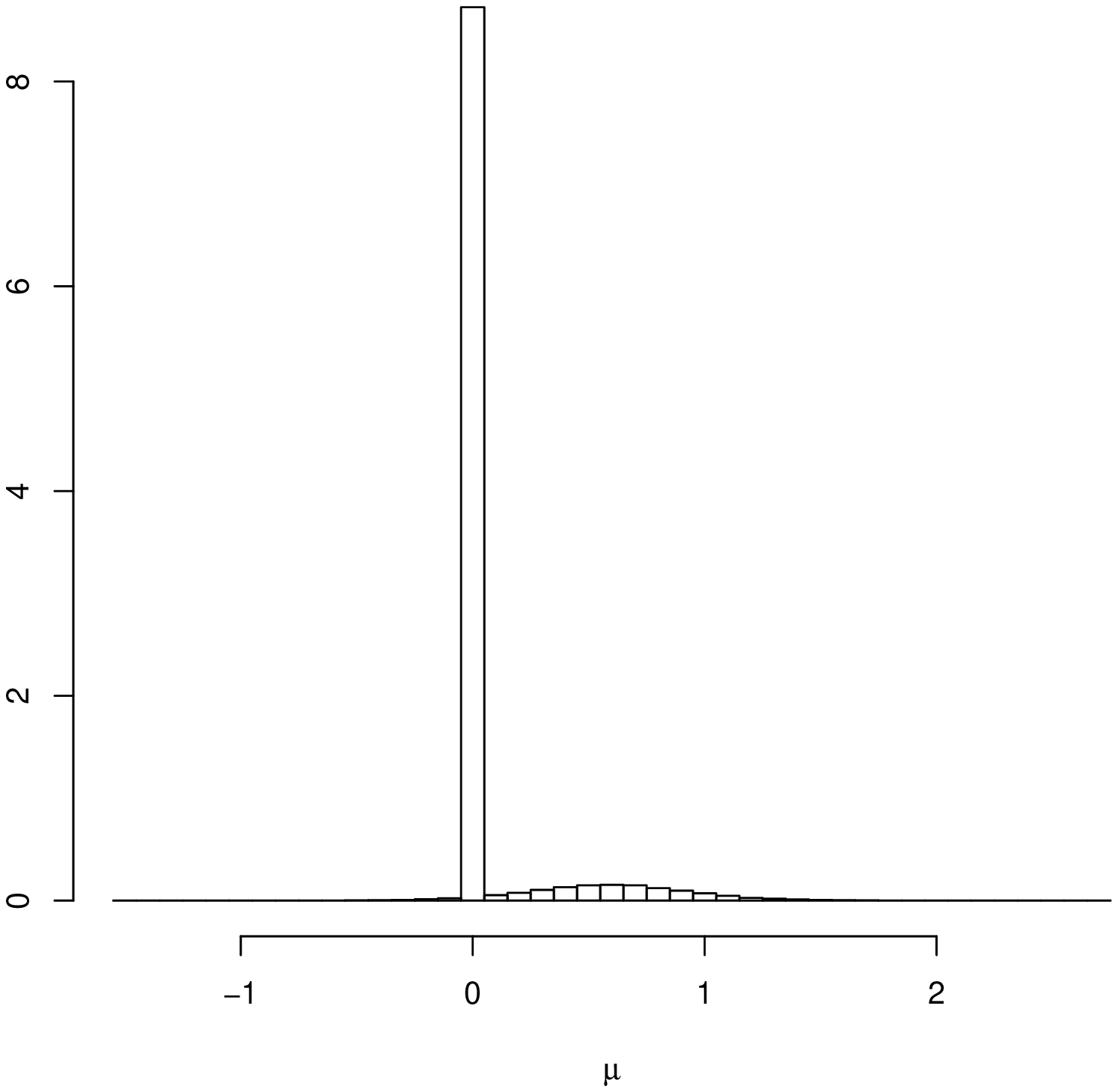,height=2.6in,width=2.8in}
\qquad
\psfig{figure=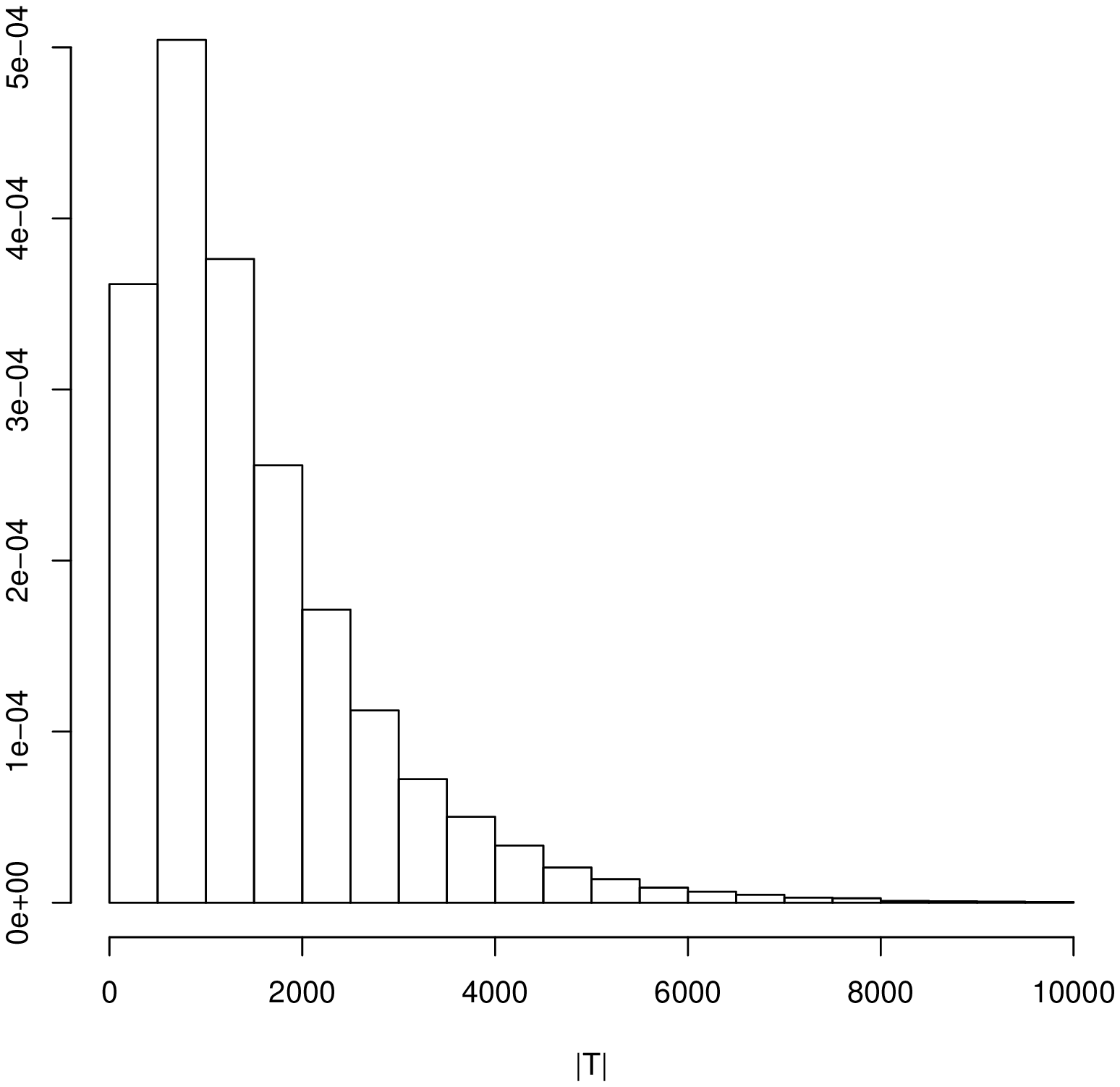,height=2.6in,width=2.8in} \\
{\small{(a)}}\qquad \qquad \qquad \quad \qquad \qquad \qquad \qquad
\qquad {\small{(b)}}
\end{center}
\end{figure}

The resulting values of $\mu$ marginalized from a perfect class coupler simulation of
$\pi(\mu,v)$ are shown in Figure (\ref{fig:page11mus}a). The
estimated posterior probability that $\mu=0$, based on $100,000$ independent draws,
 $0.86907$ which gives an approximate
95\% confidence interval for this posterior probability as
$(0.86694,0.87120)$.

This gives positive, but not what is conventionally viewed as strong
evidence (\cite{gotraft2004}, \cite{kassraft95}), for the null
hypothesis, $H_{0}: \mu = 0$.

For comparison, the Gottardo and Raftery \cite{gotraft2004} paper gives
three different estimates and standard deviations for the posterior
probability that $\mu = 0$. They are
$$
0.858 \, (0.0054), \,\,\, 0.870 \, (0.0031), \,\,\, \mbox{and} \,\,\,
0.865 \, (0.0036).
$$
The first and second estimates are from forward Metropolis-Hastings
algorithms with different candidate distributions. The third estimate is
from a Gibbs sampler. All are based on $10,000$ dependent draws from a
single Markov chain collected after a burn-in period of $1,000$ time
steps. Although they are close to our estimate, the length of the burn
in period seems to have been arbitrarily chosen and convergence is not
assured. For our results, we have chosen to produce independent draws
for $\mu$ as allowed by the speed of the class coupler. The mean
backward coupling time in our $100,000$ draws was $1502.6$ with
a minimum of $6$ and a maximum of $8223$. A histogram of these
backward
coupling times  is shown in Figure
(\ref{fig:page11mus}b).

\vspace{0.2in}
{\bf Results: Simulation Two:}

\begin{figure}[htp]
\caption{Simulation Results for \Section{simpleregression}, Simulation
  Two a) Histogram of 100,000 Values of $\mu$ Drawn from
the Posterior $\pi(\mu,v)$, b) Histogram of Corresponding
Backward Coupling Times}
\label{fig:page11mus2}
\begin{center}
\
\psfig{figure=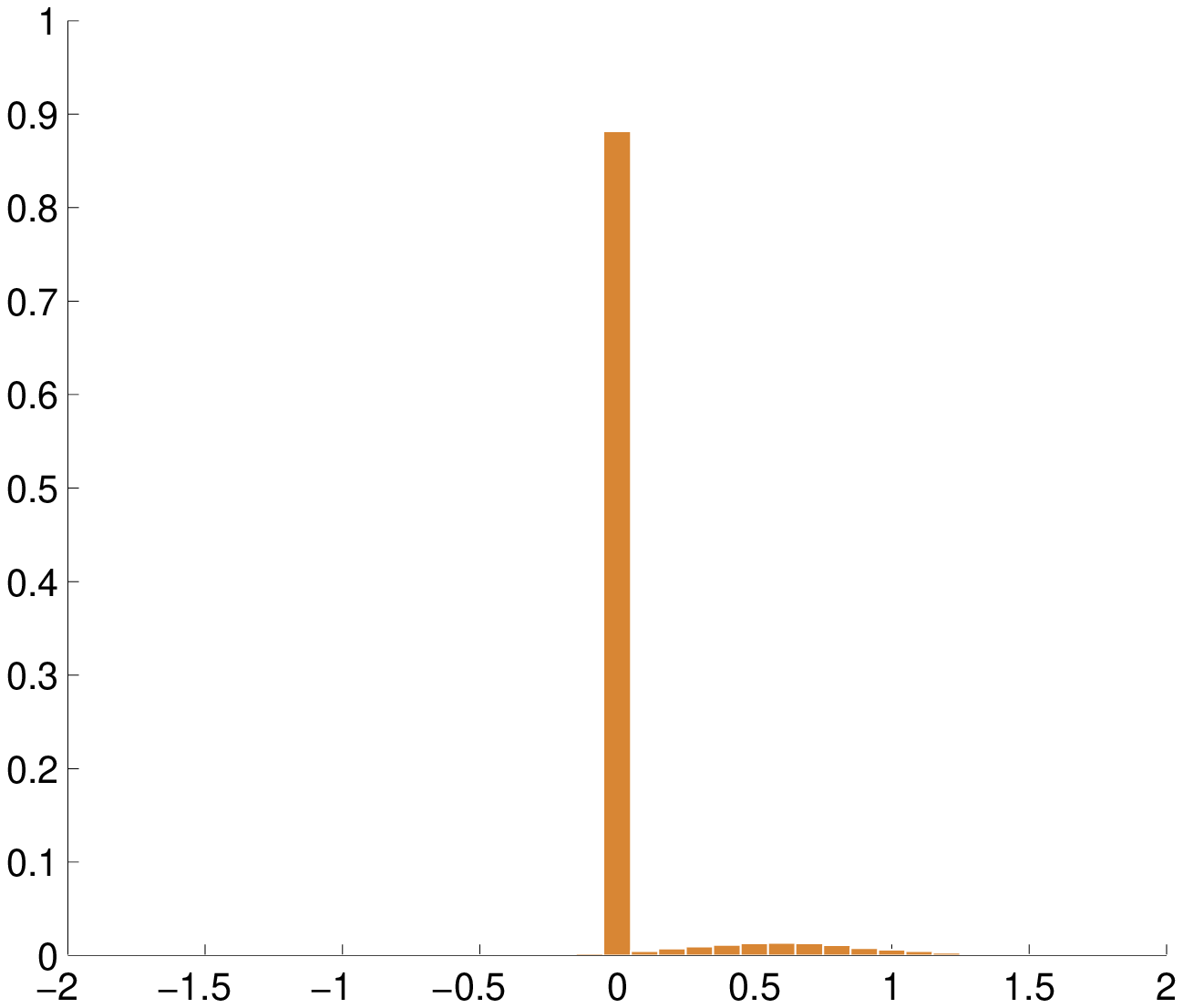,height=2.5in,width=2.8in}
\qquad
\psfig{figure=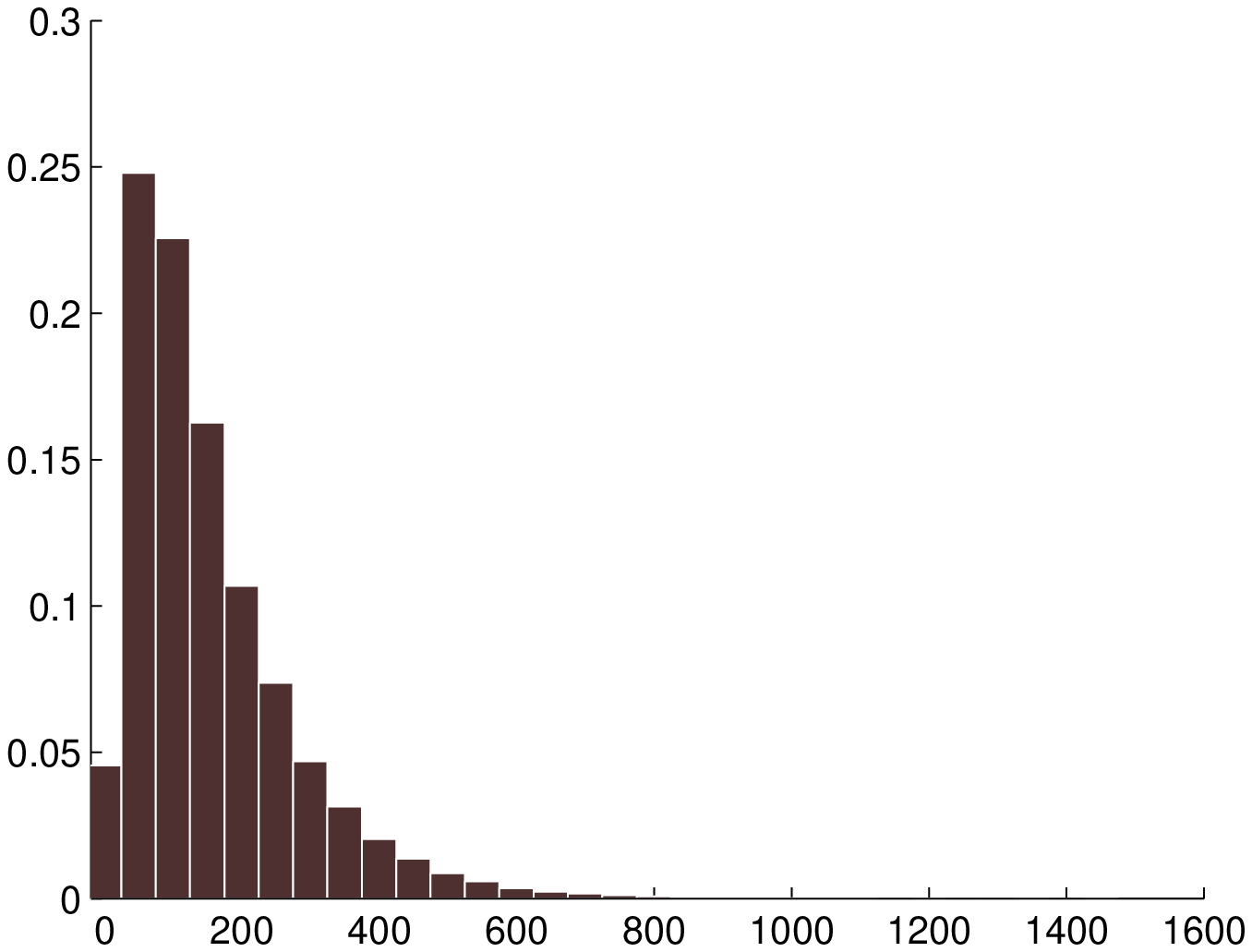,height=2.5in,width=2.8in}
{\small{(a)}}\qquad \qquad \qquad \qquad \qquad \qquad \qquad \qquad
\qquad \qquad {\small{(b)}}
\end{center}
\end{figure}

For this simulation, the data, model, and parameters are the same as in
Simulation One with the exception that $k_{2}$ has been changed from
$0.05$ to $1$. This makes the variance for the error term $\epsilon_{j}$
much smaller, so it is not unexpected that we see a much smaller
backward coupling time. Based on $100,000$ draws, the estimated
posterior probability for the mean zero model is now $0.87855$. The mean
backward coupling time is $153.13$ with a minimum value of $1469$, and a
maximum value of $3$. Histograms are given in Figures
(\ref{fig:page11mus2}a) and (\ref{fig:page11mus2}b).

\subsection{A Two-Sample Problems}
\slabel{twosample}
We now consider the model 
$$
Y_{ij} = \mu_{i} + \epsilon_{ij}, \qquad i=1,2, \,\,\, \mbox{and} \,\,\, j=1,2,\ldots, n_{i},
$$
where the $\epsilon_{ij}$ are all independent and 
$$
\epsilon_{ij} \sim N(0, v_{i}) \qquad \mbox{for} \,\,\, i=1,2,
$$
though other error distributions may easily be substituted.

Suppose that we wish to test 
$$
H_{0} : \mu_{1} = \mu_{2}
$$
versus
$$
H_{1} : \mu_{1} \ne \mu_{2},
$$
in the cases where
\begin{enumerate}
\item $v_{1}$ and $v_{2}$ are fixed and known,
\item $v_{1} = v_{2}$ are unknown and given an inverse gamma prior, and
\item $v_{1} \ne v_{2}$ are unknown and given independent inverse
  gamma priors.
\end{enumerate}

We consider a mixture prior density for $\mu=(\mu_{1},\mu_{2})$ that
allows for the two components to possibly be equal:
\beq
\elabel{2parammixture}
f(\mu_{1},\mu_{2}|p,\theta_{1},\theta_{2}) = p \cdot \delta_{0}(\mu_{1}-\mu_{2})
\cdot f_{1}(\mu_{1}|\theta_{1}) + (1-p) \cdot f_{2}
(\mu_{1},\mu_{2}|\theta_{2})
\eeq
with known hyperparameters $p$, $\theta_{1}$, and $\theta_{2}$.

The above cases can all be handled by our class coupler algorithm.

\vspace{0.1in}
{\bf Case 1: $v_{1}$ and $v_{2}$ are fixed and known}

We use the transition candidate density
$$
q(\mu,\mu^{\prime})=
q((\mu_{1},\mu_{2}),(\mu_{1}^{\prime},\mu_{2}^{\prime})) = \left\{
\begin{array}{lcl}
f_{2}(\mu_{1}^{\prime},\mu_{2}^{\prime}|\theta_{2}) &, & \mbox{if} \,\,\, \mu_{1} = \mu_{2}\\
\\
\delta_{0}(\mu_{1}^{\prime}-\mu_{2}^{\prime}) \cdot f_{1}(\mu_{1}^{\prime}|\theta_{1}) &, & \mbox{if} \,\,\, \mu_{1} \ne \mu_{2}.\\
\end{array}
\right.
$$
Then the acceptance probability ratio is
$$
r(\mu,\mu^{\prime}) = 
\frac{\pi(\mu^{\prime})q(\mu^{\prime},\mu)}{\pi(\mu) q(\mu,\mu^{\prime})} =
\frac{f(\vec{y}|\mu^{\prime})
f(\mu^{\prime}|p,\theta_{1},\theta_{2})
q(\mu^{\prime},\mu)}{f(\vec{y}|\mu) f(\mu|p,\theta_{1},\theta_{2})
q(\mu,\mu^{\prime})}.
$$
If we define 
$$
\mbox{class I} = \{(\mu_{1},\mu_{2}): \mu_{1}=\mu_{2}\}
$$
and
$$
\mbox{class II} = \{(\mu_{1},\mu_{2}): \mu_{1} \ne \mu_{2}\},
$$
then we can be assured that all class I points $\mu=(\mu_{1},\mu_{2})$ will accept the class II candidate $\mu^{\prime} = (\mu_{1}^{\prime},\mu_{2}^{\prime})$ with probability
$$
\begin{array}{lcl}
\min_{\mu} \, r(\mu,\mu^{\prime}) & = & \min_{\mu} \, \frac{f(\vec{y}|\mu^{\prime}) (1-p) \cdot f_{2}(\mu_{1}^{\prime},\mu_{2}^{\prime}|\theta_{2}) \cdot \delta_{0}(\mu_{1}-\mu_{2}) \cdot f_{1} (\mu_{1}|\theta_{1})}{f(\vec{y}|\mu) \cdot p \cdot \delta_{0}(\mu_{1}-\mu_{2}) f_{1}(\mu_{1}|\theta_{1})\cdot f_{2}(\mu_{1},\mu_{2}|\theta_{2})}\\
\\
&=& \min_{\mu} \, \frac{1-p}{p} \cdot \frac{f(\vec{y}|\mu^{\prime})}{f(\vec{y}|\mu)}.
\end{array}
$$

The minimum occurs when $\mu$ is the restricted MLE, 
$$
\hat{\mu}_{0} =  \left( \frac{1}{n_{1}+n_{2}} \sum_{i=1}^{2} \sum_{j=1}^{n_{i}} Y_{ij},  \frac{1}{n_{1}+n_{2}} \sum_{i=1}^{2} \sum_{j=1}^{n_{i}} Y_{ij} \right).
$$

Similar calculations show that all class II points will accept the class I candidate $\mu$ with probability $r(\hat{\mu},\mu)$
where $\hat{\mu}$ is the unrestricted MLE given by
$$
\hat{\mu} = \left( \frac{1}{n_{1}} \sum_{i=1}^{n_{1}} Y_{1i}, \frac{1}{n_{2}} \sum_{i=1}^{n_{2}} Y_{2i}  \right).
$$

Therefore, we may run the same algorithm as in \Section{simpleregression}.

\vspace{0.2in}
{\bf Case 2: $v_{1} = v_{2}$ are unknown and given
  an inverse gamma prior}

We will use $v$ to denote the common value of $v_{1}$ and $v_{2}$. We
assume 
an inverse gamma prior:
$$
v \sim IG(k_{1},k_{2})
$$
where $k_{1}$ and $k_{2}$ are known hyperparameters, and we assume that
the $\mu = (\mu_{1},\mu_{2})$ and $v$ are a priori independent.

We choose the transition candidate density
$$
\begin{array}{lcl}
q((\mu,v),(\mu^{\prime},v^{\prime}))&=&
q((\mu_{1},\mu_{2},v),(\mu_{1}^{\prime},\mu_{2}^{\prime},v^{\prime}))\\
\\
& =& \left\{
\begin{array}{lcl}
f_{2}(\mu_{1}^{\prime},\mu_{2}^{\prime}|\theta_{2}) \cdot IG(v;k_{1},k_{2}) &, & \mbox{if} \,\,\, \mu_{1} = \mu_{2}\\
\\
\delta_{0}(\mu_{1}^{\prime}-\mu_{2}^{\prime}) \cdot f_{1}(\mu_{1}^{\prime};\theta_{\mu}) \cdot IG(v^{\prime};k_{1},k_{2}) &, & \mbox{if} 
\,\,\,\
 \mu_{1} \ne \mu_{2}.\\
\end{array}
\right.
\end{array}
$$

The acceptance probability ratio is
$$
\begin{array}{lcl}
r((\mu,v),(\mu^{\prime},v^{\prime})) &=&
\frac{\pi(\mu^{\prime},v^{\prime })q((\mu^{\prime},v^{\prime}),(\mu,v))}{\pi(\mu,v) q((\mu,v),(\mu^{\prime},v^{\prime}))} \\
\\
&=&
\frac{f(\vec{y}|\mu^{\prime},v^{\prime})
f(\mu^{\prime}|p,\theta_{1},\theta_{2}) IG(v^{\prime};k_{1},k_{2})
q((\mu^{\prime},v^{\prime}),(\mu,v))}{f(\vec{y}|\mu,v) f(\mu|p,\theta_{1},\theta_{2}) IG(v;k_{1},k_{2})
q((\mu,v),(\mu^{\prime},v^{\prime}))}.
\end{array}
$$

If we define
$$
\mbox{class I} = \{(\mu_{1},\mu_{2},v): \mu_{1}=\mu_{2}\}
$$
and
$$
\mbox{class II} = \{(\mu_{1},\mu_{2},v): \mu_{1} \ne \mu_{2}\},
$$
then we can be assured that all class I points
$(\mu,v)=(\mu_{1},\mu_{2},v)$ 
will accept the class II candidate 
$(\mu^{\prime},v^{\prime}) = (\mu_{1}^{\prime},\mu_{2}^{\prime},v^{\prime})$ with probability
$$
r((\hat{\mu}_{0},\hat{v}_{0}),(\mu^{\prime},v^{\prime}))
$$
where $\hat{\mu}_{0}$ and $\hat{v}_{0}$ are the restricted MLEs
$$
\hat{\mu}_{0} =  \left( \overline{\overline{Y}}, \overline{\overline{Y}} \right)=\left( \frac{1}{n_{1}+n_{2}} \sum_{i=1}^{2} \sum_{j=1}^{n_{i}} Y_{ij},  \frac{1}{n_{1}+n_{2}} \sum_{i=1}^{2} \sum_{j=1}^{n_{i}} Y_{ij} \right) 
$$
and
$$
\hat{v}_{0} = \frac{1}{n_{1}+n_{2}} \sum_{i=1}^{2} \sum_{j=1}^{n_{i}} (Y_{ij}-\overline{\overline{Y}})^{2}.
$$
Similarly, all class II points $(\mu,v)$
will accept the class I candidate
$(\mu^{\prime},v^{\prime})$ 
with probability
$$
r((\hat{\mu},\hat{v}),(\mu^{\prime},v^{\prime}))
$$
where $\hat{\mu}$ and $\hat{v}$ are the unrestricted MLEs
$$
\hat{\mu} = (\overline{Y}_{1}, \overline{Y}_{2}) = \left( \frac{1}{n_{1}} \sum_{j=1}^{n_{1}} Y_{1j}, 
\frac{1}{n_{2}} \sum_{j=1}^{n_{2}} Y_{2j} \right) 
$$
and
$$
\hat{v}^{2} =  \frac{1}{n_{1}+n_{2}} \left[ \sum_{i=1}^{n_{1}} (Y_{1i}-\overline{Y}_{1})^{2} + \sum_{j=1}^{n_{2}} (Y_{2j}-\overline{Y}_{2})^{2} \right].
$$

\vspace{0.2in}
{\bf Case 3: $v_{1} \ne v_{2}$ are unknown and given independent inverse gamma priors}

The prior for $(\mu,v) = (\mu_{1},\mu_{2},v_{1},v_{2})$ is
$$
f(\mu|p,\theta_{1},\theta_{2}) \cdot IG(\sigma_{1}^{2};k_{11},k_{12}) \cdot IG(\sigma_{2}^{2};k_{21},k_{22})
$$
where $f_{\mu}(\mu;\theta_{\mu})$ is from (\ref{e:2parammixture}) and
$k_{11}, k_{12}, k_{21}$, and $k_{22}$ are known hyperparameters.

We choose the transition candidate density
$$
q((\mu,v),(\mu^{\prime},v^{\prime}))=
q((\mu_{1},\mu_{2},v_{1},v_{2}),(\mu_{1}^{\prime},\mu_{2}^{\prime},v_{1}^{\prime},v_{2}^{\prime}))
$$

$$
 = \left\{
\begin{array}{lcl}
f_{2}(\mu_{1}^{\prime},\mu_{2}^{\prime};\theta_{2}) \cdot IG(v_{1}^{\prime};k_{11},k_{12}) \cdot IG(v_{2}^{\prime};k_{21},k_{22})  &, \
& \mbox{if} \,\,\, \mu_{1} = \mu_{2}\\
\\
\delta_{0}(\mu_{1}^{\prime}-\mu_{2}^{\prime}) \cdot f_{1}(\mu_{1}^{\prime}|\theta_{1}) \cdot 
IG(v_{1}^{\prime};k_{11},k_{12}) \cdot IG(v_{2}^{\prime};k_{21},k_{22}) &, & \mbox{if}
\,\,\,\
 \mu_{1} \ne \mu_{2}.\\
\end{array}
\right.
$$

The acceptance probability ratio is
$$
\begin{array}{lcl}
r((\mu,v),(\mu^{\prime},v^{\prime})) &=&
\frac{\pi(\mu^{\prime},v^{\prime})q((\mu^{\prime},v^{\prime}),
(\mu,v))}{\pi(\mu,v) q((\mu,v),(\mu^{\prime},
v^{\prime}))} \\
\\
&=&
\frac{f(\vec{y}|\mu^{\prime},v^{\prime})
f(\mu^{\prime}|p,\theta_{1},\theta_{2}) IG(v_{1}^{\prime};k_{11},k_{12}) 
IG(v_{2}^{\prime};k_{21},k_{22})
q((\mu^{\prime},v^{\prime}),(\mu,v))}{f
(\vec{y}|\mu,v) f(\mu|p,\theta_{1},\theta_{2}) 
IG(v_{1};k_{1},k_{2})
IG(v_{2};k_{21},k_{22})
q((\mu,v),(\mu^{\prime},v^{\prime}))}.
\end{array}
$$
Defining class I and class II points in the same way as in Case 2 above, we are assured that all class I points $(\mu,v) = (\mu_{1},\mu_{2},v_{1},v_{2})$ will accept the class II candidate $(\mu^{\prime},v^{\prime}) = 
(\mu_{1}^{\prime},\mu_{2}^{\prime},v_{1}^{\prime},v_{2}^{\prime})$ with probability
$$
r((\hat{\mu}_{0},\hat{v}_{0}),(\mu^{\prime} ,v^{\prime})
$$
where $\hat{\mu}_{0}$ and $\hat{v}_{0}$ are the restricted MLEs
$$
\hat{\mu}_{0} = \left( \overline{\overline{Y}},\overline{\overline{Y}} \right)=\left( \frac{1}{n_{1}+n_{2}} \sum_{i=1}^{2} \sum_{j=1}^{n_{i}}\
 Y_{ij},  \frac{1}{n_{1}+n_{2}} \sum_{i=1}^{2} \sum_{j=1}^{n_{i}} Y_{ij} \right)
$$
and
$$
\hat{v}_{0} = \left( \frac{1}{n_{1}} \sum_{i=1}^{n_{1}} (Y_{1i}-\overline{\overline{Y}})^{2}, \frac{1}{n_{2}} \sum_{j=1}^{n_{2}} (Y_{2j}-\overline{\overline{Y}})^{2} \right).
$$
Similarly, all class II points $(\mu,v) = (\mu_{1},\mu_{2},v_{1},v_{2})$ will accept the class I candidate $(\mu^{\prime},v^{\prime}) =(\mu_{1}^{\prime},\mu_{2}^{\prime},v_{1}^{\prime},v_{2}^{\prime})$ with probability
$$
r((\hat{\mu},\hat{v}),(\mu^{\prime},v^{\prime}))
$$
where $\hat{\mu}$ and $\hat{v}$ are the unrestricted MLEs
$$
\hat{\mu} = (\overline{Y}_{1}, \overline{Y}_{2}) = \left( \frac{1}{n_{1}} 
\sum_{j=1}^{n_{1}} Y_{1j},
\frac{1}{n_{2}} \sum_{j=1}^{n_{2}} Y_{2j} \right)
$$
and
$$
\hat{v} =  \left( \frac{1}{n_{1}} \sum_{i=1}^{n_{1}} (Y_{1i}-\overline{Y}_{1})^{2}, \frac{1}{n_{2}} \sum_{j=1}^{n_{2}} (Y_{2j}-\overline{Y}_{2})^{2}
\right).
$$

%
\bibliographystyle{acm}

\end{document}